\documentclass[11pt,a4paper]{article}
\usepackage{amsmath,amssymb,mathrsfs}
\usepackage[dvips]{graphicx}
\usepackage{epsfig}

\def\un#1{\relax\ifmmode\@@underline#1\else
        $\@@underline{\hbox{#1}}$\relax\fi}

\let\du=\du                     

\def\l{\lambda}
\def\m{\mu}
\def\n{\nu}

\def\L{\Lambda}

\def\bo{{\raise-.3ex\hbox{\large$\Box$}}}               
\def\TH{{\raise.2ex\hbox{$\displaystyle \bigodot$}\mskip-4.7mu \llap H \;}}
\def\face{{\raise.2ex\hbox{$\displaystyle \bigodot$}\mskip-2.2mu \llap {$\ddot
        \smile$}}}                                      

\def\abs#1{\left| #1\right|}                    
\def\leftrightarrowfill{$\mathsurround=0pt \mathord\leftarrow \mkern-6mu
        \cleaders\hbox{$\mkern-2mu \mathord- \mkern-2mu$}\hfill
        \mkern-6mu \mathord\rightarrow$}
\def\dvec#1{\vbox{\ialign{##\crcr
        \leftrightarrowfill\crcr\noalign{\kern-1pt\nointerlineskip}
        $\hfil\displaystyle{#1}\hfil$\crcr}}}           

\def\sfrac#1#2{{\vphantom1\smash{\lower.5ex\hbox{\small$#1$}}\over
        \vphantom1\smash{\raise.4ex\hbox{\small$#2$}}}} 
\def\bfrac#1#2{{\vphantom1\smash{\lower.5ex\hbox{$#1$}}\over
        \vphantom1\smash{\raise.3ex\hbox{$#2$}}}}       
\def\afrac#1#2{{\vphantom1\smash{\lower.5ex\hbox{$#1$}}\over#2}}    

\def\[{\lfloor{\hskip 0.35pt}\!\!\!\lceil}
\def\]{\rfloor{\hskip 0.35pt}\!\!\!\rceil}

\def\du#1#2{_{#1}{}^{#2}}

\def\un{\underline}
\def\fracmm#1#2{{{#1}\over{#2}}}

\def\low#1{{\raise -3pt\hbox{${\hskip 0.75pt}\!_{#1}$}}}

\newskip\humongous \humongous=0pt plus 1000pt minus 1000pt
\def\caja{\mathsurround=0pt}
\def\eqalign#1{\,\vcenter{\openup2\jot \caja
        \ialign{\strut \hfil$\displaystyle{##}$&$
        \displaystyle{{}##}$\hfil\crcr#1\crcr}}\,}
\newif\ifdtup

\newcommand{\be}{\begin{equation}}
\newcommand{\ee}{\end{equation}}
\newcommand{\nbe}{\begin{equation*}}
\newcommand{\nee}{\end{equation*}}

\newcommand{\lb}{\label}

\def\lessim{\lower0.6ex\hbox{$\,$\vbox{\offinterlineskip\hbox{$<$}\vskip1pt\hbox{$\sim$}}$\,$}}
\def\grtsim{\lower0.6ex\hbox{$\,$\vbox{\offinterlineskip\hbox{$>$}\vskip1pt\hbox{$\sim$}}$\,$}}

\begin{document}

\begin{titlepage}

\begin{center}

revised version \hfill IPMU14-0005\\

\noindent
\vskip2.0cm
{\Large \bf 

On the Higgs-like Quintessence for Dark Energy

}

\vglue.3in

{\large
Sergei V. Ketov~${}^{a,b,c}$ and Natsuki Watanabe~${}^a$ 
}

\vglue.1in

{\em
${}^a$~Department of Physics, Tokyo Metropolitan University \\
Minami-ohsawa 1-1, Hachioji-shi, Tokyo 192-0397, Japan \\
${}^b$~Kavli Institute for the Physics and Mathematics of the Universe (IPMU)
\\The University of Tokyo, Chiba 277-8568, Japan \\
${}^c$~Institute of Physics and Technology, Tomsk Polytechnic University\\
30 Lenin Ave.,Tomsk 634050, Russian Federation \\
}

\vglue.1in
ketov@tmu.ac.jp, watanabe-natsuki1@tmu.ac.jp

\end{center}

\vglue.3in

\begin{center}
{\large\bf Abstract}
\end{center}

We propose a dynamical (quintessence) model of dark energy in the current universe with a renormalizable 
(Higgs-like) scalar potential. We prove the viability of our model (after fine tuning) for the certain range of the 
average scalar curvature values, and study the cosmological signatures distinguishing our model from the 
standard description of dark energy in terms of a cosmological constant.

\end{titlepage}


\section{Introduction}\label{sec:Intro}
The dynamical dark energy models of modified gravity are usually constructed in the framework of $f(R)$ gravity --- see e.g., Refs.~\cite{svrev,tsu,norev,clrev,myrev} for some reviews with many references therein, and Refs.~\cite{ab,hs,stard} for some viable proposals to the $f(R)$-function, that are directly related to this paper. 
 The current status of the $f(R)$ gravity theories is phenomenological (or macroscopic) and is truly non-perturbative (or non-linear). The $f(R)$-functions are chosen {\it ad hoc}, in order to satisfy the existing phenomenological constraints coming from the Newtonian limit, classical and quantum stability, the Solar System tests and the cosmological tests. The usual treatment includes rewriting an $f(R)$-gravity model into the classically equivalent scalar-tensor gravity (or quintessence) by the Legendre-Weyl transform from the Jordan frame to the Einstein frame \cite{wtt}, and then applying the standard cosmology in terms of the dual (quintessence) scalar potential (see also Sec.~2). For instance, the 5th force (due to the quintessence scalar) is screened at high matter density (like that of the Solar system) by the Chameleon effect \cite{hw}. 

The scalar potentials, originating from the $f(R)$ gravity functions of Refs.~\cite{ab,hs,stard}, are very 
complicated and non-renormalizable. In this Letter we physically motivate a shape of the quintessence scalar potential, and then use the {\it inverse} Legendre-Weyl transformation, in order to determine yet another $f(R)$ function that is suitable for describing the present dark energy in the Universe and is related to a renormalizable quintessence scalar potential.

The paper is organized as follows. In Sec.~2 we review the Legendre-Weyl transformation and give its inverse form. 
In Sec.~3 we propose the scalar potential of the Higgs-type with the {\it Uplifted-Double-Well} (UDW) shape, and
find the corresponding $f(R)$ function. Further, we demonstrate that our model is viable, being close to the standard Einstein gravity with a (positive) cosmological constant $\Lambda$. In Sec.~4 we study the modified gravity corrections (beyond the cosmological constant), in order to distinguish our model from the standard
description of the present dark energy by the $\Lambda$, by using MATHEMATICA. Sec.~5 is our conclusion.    

Throughout this paper we use the natural units, $c = \hbar = 1$, and the space-time signature 
$\eta_{\mu \nu} = {\rm diag} (1, - 1, - 1, - 1)$. The Einstein-Hilbert action with a cosmological constant $\L$ reads
\begin{equation} \lb{eh}
 S_{\rm EH} =  \frac{1}{2 \kappa^{2}} \int d^{4}x \sqrt{- g}~(-R - 2 \Lambda),
 \end{equation} 
where $R$ is the scalar curvature, $\kappa^{2} = \fracmm{1}{M_{\rm Pl}^{2}} = 1.7 \times 
10^{- 37}~{\rm GeV^{-2}}$, and $M_{\rm Pl} = ( 8 \pi G_{\rm N} )^{- 1 / 2}$ is the (reduced) Planck mass in terms
of the Newton constant $G_{\rm N}$. In our notation here, the cosmological constant $\L$ is positive in a {\it 
de-Sitter} (dS) space-time (like the present Universe).

A generic $f(R)$-gravity action reads
\begin{equation} \lb{fg}
  S_{f} = \frac{1}{2 \kappa^{2}} \int d^{4}x \sqrt{- g}~ f(R)
\end{equation} 
with a function $f(R)$ of the scalar curvature $R$. So that, in our notation, a de-Sitter space-time has a negative
scalar curvature. The $f(R)$ gravity can be considered as the modified gravity theory extending the Einstein gravity 
theory with a cosmological constant (= static dark energy) defined by Eq.~(\ref{eh}) to the gravitational theory with a dynamical dark energy.

An expansion of the $f(R)$ function in power series of $R$ near a de-Sitter vacuum gives rise to the modified
gravity corrections to the static dark energy. For example, the simplest model with
\begin{equation} \lb{star}
  f(R)  = -R + \frac{1}{6 M^{2}} R^2
\end{equation} 
is known in cosmology as the Starobinsky model \cite{star}. It is well suitable for describing the early universe
inflation, with the inflaton mass $M$ (see e.g., Ref.~\cite{kkw}). Hence, inflation can also be considered as a (primordial) dark energy.

\section{The Legendre-Weyl transform and its inverse}\label{sec:Setup}

The $f(R)$ gravity action (\ref{fg}) is classically equivalent to 
\begin{equation} \lb{eq}
 S[g_{\m\n},\chi] = \frac{1}{2 \kappa^{2}} \int d^{4}x \sqrt{- g}~ \left[ f'(\chi) (R - \chi) + f(\chi) \right]
\end{equation}
with the real scalar field $\chi$, provided that $f''\neq 0$ that we always assume.  Here the
primes denote the differentiation. The equivalence is easy to verify because the $\chi$-field equation implies $\chi=R$.

The factor $f'$ in front of the $R$ in eq.~(\ref{eq}) can be eliminated by a Weyl transformation of 
metric $g_{\m\n}$, so that one can transform the action (\ref{eq}) into the action of the scalar field 
$\chi$ miminally coupled to the Einstein gravity and having the scalar potential \cite{wtt}  
\be \lb{spot}
 V = \dfrac{\chi f'(\chi) - f(\chi)}{2 \kappa^{2} f'(\chi)^{2}}~~.
\ee
The kinetic term of $\chi$ becomes canonically normalized after the field refedinition
\begin{equation} \lb{fred}
  f'(\chi) = - \exp \left( - \sqrt{\frac{2}{3}} \kappa \phi \right)
\end{equation}
in terms of the new scalar field $\phi$. As a result, the action $S[g_{\m\n},\chi(\phi)]$  takes the standard
quintessence form \cite{wein}.

The classical and quantum stability conditions (in our notation) are given by \cite{tsu,myrev}
\begin{equation} \lb{stab}
 f'(\chi) < 0 \quad {\rm and} \quad f''(\chi)> 0~, 
\end{equation}
respectively. The first condition ensures the existence of a solution to Eq.~(\ref{fred}).

The mass dimensions of various quantities are given by
\begin{equation} \lb{dim}
  [\kappa] = - 1, \quad
  [\phi] = 1, \quad
  [f] = [\chi] = [R] = [\Lambda] = 2, \quad
  [V] = 4.
\end{equation}

Differentiating the scalar potential $V$ in Eq.~(\ref{spot}) with respect to $\phi$ yields 
\begin{equation} \lb{diff1}
    \frac{d V}{d \phi}   = \frac{d V}{d \chi} \frac{d \chi}{d \phi} 
      = \frac{1}{2 \kappa^{2}} \left[ \frac{\chi f'' + f' - f'}{f'^{2}} - 2 \frac{\chi f' - f}{f'^{3}} f'' \right] 
\frac{d \chi}{d \phi}~~,
\end{equation}
where we have
\begin{equation} \lb{diff2}
  \frac{d \chi}{d \phi}
    = \frac{d \chi}{d f'} \frac{ d f'}{d \phi}
    = \frac{d f'}{d \phi} \left/ \frac{d f'}{d\chi} \right.
    = - \sqrt{\frac{2}{3}} \kappa \frac{f'}{f''}
\end{equation}
It implies that
\begin{equation} \lb{derv}
  \frac{d V}{d \phi} = \frac{\chi f' - 2 f}{\sqrt{6} \kappa f'^{2}}~~.
\end{equation}

Combining Eqs.~(\ref{spot}) and (\ref{derv})  yields $R$ and $f$ in terms of the scalar potential $V$ as follows:
\begin{align} \lb{inv}
  & R = - \left( - \sqrt{6} \kappa \frac{d V}{d \phi} + 4 \kappa^{2} V \right) \exp \left( - \sqrt{\frac{2}{3}} \kappa \phi \right),  \\
  & f = \left( - \sqrt{6} \kappa \frac{d V}{d \phi} + 2 \kappa^{2} V \right) \exp \left( - 2 \sqrt{\frac{2}{3}} \kappa \phi \right).
\end{align}
These two equations define the function $f(R)$ in the parametric form in terms of a given scalar potential $V(\phi)$. It is the inverse transformation against defining the scalar potential $V(\phi)$ in terms of a given 
$f(R)$ function, according to Eqs.~(\ref{spot}) and (\ref{fred}).
 
\section{The UDW Scalar Potential}\label{sec:Potential}

A choice of the quintessence scalar potential $V(\phi)$ is usually dictated by desired phenomenology without imposing formal constraints.  It is in the striking difference with the Standard Model of elementary particles whose Higgs scalar potential of the Double-Well (DW) shape is severely constrained by renormalizability to a {\it quartic} function of the Higgs field. In fact, it is one of the basic reasons for high predictability of the Higgs-based physics of elementary particles!

Since we are interested in the existence of a de Sitter vacuum for describing dark energy in $f(R)$ gravity, we uplift one of the two DW-vacua of the Higgs scalar potential, while keeping another one to be a Minkowski  (flat) space-time, as in Fig.~1. The potential barrier between those two vacua has to be high enough, in order to be consistent with the long lifetime of our Universe. It implies that the de Sitter vacuum is meta-stable and may be considered as a
``false" vacuum against the ``true" Minkowski vacuum.

We write down such {\it Uplifted-Double-Well} (UDW) quartic scalar potential  in the form
\cite{kw}
\begin{equation} \lb{udwp}
  V_{\rm UDW}(y) =
    \frac{\lambda}{2 \kappa^{2}} \left\{ \left[ ( y - y_{0} )^{2} - u^{2} \right]^{2}
      + \mu^{2} \left[ ( y - y_{0} ) - u \right]^{2} \right\},
\end{equation}
where we have introduced the field
\begin{equation} \lb{yf}
  y = \sqrt{\frac{2}{3}} \kappa \phi
\end{equation}
and four real parameters $y_{0}$ , $u$, $\lambda$ and $\mu$ of (mass) dimension
\begin{equation} \lb{pdim}
  [\lambda] = 2, \quad  [y] = [y_{0}] = [u] = [\mu] = 0.
\end{equation}
Amongst those parameters, the $\l$ sets the dimensional scale, the $y_0$ is physically irrelevant (it can be changed by a shift of the field $y$), so that the shape of the UDW potential is determined by only two 
dimensionless parameters $(u,\mu)$.  

The parametrization of the Higgs-type scalar potential used in Eq.~(\ref{udwp}) is convenient for several reasons. First, it is a generic parametrization compatible with the UDW shape. Second, it
allows us to have the mass $m$ (of the quintessence scalar), the cosmological constant $\Lambda$
in the dS vacuum, and the height $h$ of the barrier between the dS and Minkowski vacua to be fully independent (see below). Third, the scalar potential (\ref{udwp}) admits a supersymmetric extension of the quintessence theory, as was already demonstrated in Ref.~\cite{kw}, by using a particular (O'Raifertaigh-type) model of spontaneous supersymmetry breaking. Fourth, it appears to be easy to find the vacua of the scalar potental Eq.~(\ref{udwp})
analytically.

The UDW scalar potential (\ref{udwp}) has three extrema at
\be \lb{3ext}
y_{\rm c} = y_{0} + u  \quad {\rm and} \quad 
 y_{\pm} = y_{0} + \frac{1}{2} \left( - u \pm \sqrt{u^{2} - 2 \mu^{2}} \right)~,
\ee
where  $y_{-}$ and $y_{\rm c}$ are the positions of the de Sitter and Minkowski vacua, respectively, and
$y_{+}$ is the  position of the maximum of the potential barrier (Fig.~1).

\begin{figure}[t]
\centering
 \includegraphics[width=0.7\textwidth]{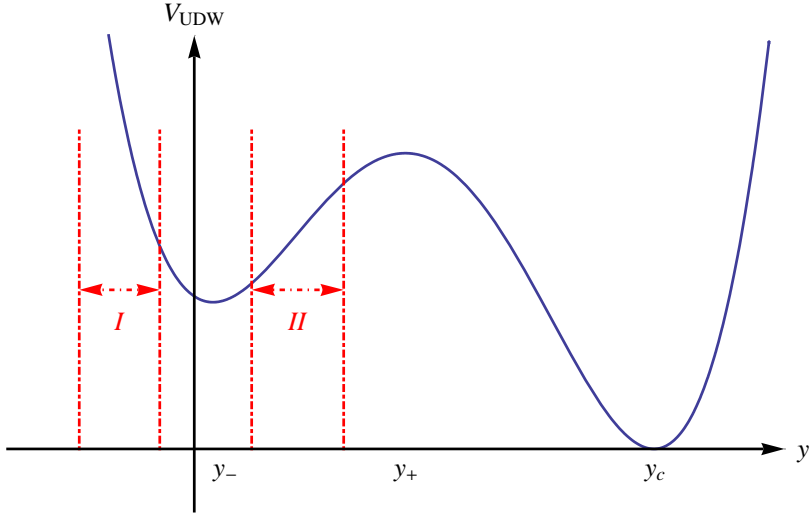}
 \caption{The position of the dS vacuum is $y_{-}$ ,  the position of the barrier (maximum) is  $y_{+}$,
and the position of the Minkowski vacuum is $y_{c}$. The regions of higher curvature near the dS vacuum 
are denoted by I and II.}
  \label{fig:UDW}
\end{figure}

The height of the barrier is given by
\begin{equation} \lb{ht}
  h=V_{\rm UDW}(y_{+}) - V_{\rm UDW}(y_{-}) = \frac{\lambda}{2 \kappa^{2}} u ( u^{2} - 2 \mu^{2})^{3 / 2}~.
\end{equation}

Differentiating the $V_{\rm UDW}$ with respect to $y$ yields
\begin{equation} \lb{dif3}
  \frac{d V_{\rm UDW}}{d y} =
    \frac{\lambda}{\kappa^{2}} \left\{ 2( y - y_{0} ) 
\left[ ( y - y_{0} )^{2} - u^{2} \right] + \mu^{2} ( y - y_{0} - u ) \right\}~,
\end{equation}
where we have 
\begin{equation} \lb{dif4}
  \frac{d V_{\rm UDW}}{d \phi}
    = \frac{d y}{d \phi} \frac{d V_{\rm UDW}}{d y}
    = \sqrt{\frac{2}{3}} \kappa \frac{d V_{\rm UDW}}{d y}~.
\end{equation}
Substituting these equations into the inverse transformation formulae of Sec.~2 yields
\begin{equation} \lb{invR}
  \eqalign{ 
    R(y) = &
      - 2 \lambda e^{- y} \left\{ y^{4} - 2 ( 2 y_{0} + 1 ) y^{3}
        + ( 6 y_{0}^{2} + 6 y_{0} - 2 u^{2} + \mu^{2} ) y^{2} \right.  \cr
          & 
        - \left[ 4 y_{0} ( y_{0}^{2} - u^{2} ) + 2 ( 3 y_{0}^{2} - u^{2} ) + \mu^{2} ( 2 y_{0} + 2 u + 1 ) \right] y  \cr
          &  \left.
        + ( y_{0}^{2} - u^{2} ) ( y_{0}^{2} + 2 y_{0} - u^{2} ) + \mu^{2} (y_{0} + u ) ( y_{0} + u + 1 ) \right\}
         \cr}
\end{equation}
and
\begin{equation} \lb{invf}
  \eqalign{
    f(y) = &
      \lambda e^{- 2 y} \left\{ y^{4} - 4 ( y_{0} + 1 ) y^{3}
        + ( 6 y_{0}^{2} + 12 y_{0} - 2 u^{2} + \mu^{2} ) y^{2} \right.  \cr
          & 
        - 2\left[ 2 y_{0} ( y_{0}^{2} - u^{2} ) + 2 (3 y_{0}^{2} - u^{2} ) + \mu^{2} (y_{0} + u + 2 ) \right] y  \cr
          & \left.
        + ( y_{0}^{2} - u^{2} ) ( y_{0}^{2} + 4 y_{0} - u^{2} )
          + \mu^{2} (y_{0} + v ) ( y_{0} + u + 2 ) \right\}~.\cr}
\end{equation}

Equations (\ref{invR})  and (\ref{invf}) determine the exact function $f(R)$ in the parametric form, which is 
suitable for numerical calculations (e.g., by using MATHEMATICA computing software). Moreover, those equations
are greatly simplified when using the very small cosmological constant and the very high potential barrier, corresponding to the present dark energy and the current age of our Universe, respectvely. Both conditions 
necessarily imply
\be \lb{approx1}
\m^2\ll u^2~.
\ee
The required smallness of the cosmological constant is then easily achieved by fine-tuning the value of $\m^2$, 
whereas the high potential barrier related to (meta)stability of our Universe in the de Sitter vacuum can be achieved by
raising the parameter $u$ to the desired value --- see Eq.~(\ref{ht}). In addition, we choose
\be \lb{approx2}
y_0 \approx u
\ee
for further simplifications. Then Eqs.~(\ref{invR}) and (\ref{invf}) are simplified to
\be \lb{invs}
\eqalign{
   R(y) = & - 2 \lambda e^{- y}
    \left[ y^{4} - 2 ( 2 u + 1 ) y^{3} + 2 u ( 2 u + 3 ) y^{2} - 4 u ( 2 u + \mu^{2} ) y + 2 \mu^{2} u ( 2 u + 1 ) \right],  \cr
   f(y) = & \lambda e^{- 2 y}
    \left[ y^{4} - 4 ( u + 1 ) y^{3} + 4 u ( u + 3 ) y^{2} - 4 u ( 2 u + \mu^{2} ) y + 4 \mu^{2} u ( u + 1 ) \right].  \cr}
\ee

Near the de Sitter vacuum, the values of $y$ are close to zero, so that we can expand both $R$ and $f$ to the order ${\cal O}(y^{2})$ as
\be \lb{exp2}
\eqalign{
   R(y) = & 4 \lambda u \left[ ( 2 u + 2 \mu^{2} u + 3 \mu^{2} ) y - \mu^{2} ( 2 u + 1 ) \right] + {\cal O}(y^{2})~,  \cr
   f(y) = & - 4 \lambda u \left[ ( 2 u + 2 \mu^{2} u + 3 \mu^{2} ) y - \mu^{2} ( u + 1 ) \right] + {\cal O}(y^{2})~. \cr}
\ee
The first equation (\ref{exp2}) can be easily inverted as
\begin{equation}
  y \approx  \frac{R + 4 \lambda \mu^{2} u ( 2 u + 1 )}{4 \lambda u ( 2 u + 2 \mu^{2} u + 3 \mu^{2} )}~.
\end{equation}
Substituting it into the second equation (\ref{exp2}) for $f$ yields
\begin{equation}  \label{fr2}
  f(R) \approx - R - 4 \lambda \mu^{2} u^{2}.
\end{equation}
It takes the form of Eq.~(\ref{eh}) with the cosmological constant
\begin{equation} \lb{cosmc}
  2\L =\abs{R_0} = 4 \lambda \mu^{2} u^{2} = 1.12 \times 10^{- 65}~[\rm eV^{2}]~.
\end{equation}

The mass $m$ of the canonically normalized  quintessence scalar $\phi$ is given by
\begin{equation} \lb{mass}
 m^2 = \frac{d^{2} V_{\rm UDW}}{d \phi^{2}} = \frac{8}{3} \lambda u^{2} \left( 1 + \frac{\mu^{2}}{4 u^{2}} \right)
    \approx \frac{8}{3} \lambda u^{2}
\end{equation}
in the dS vacuum at $y=0$, so that it is independent upon the value of the cosmological constant in Eq.~(\ref{cosmc})
indeed. Actually, we have the relation
\begin{equation} \lb{rela}
\Lambda = \fracmm{3}{4} \m^2 m^2~.
\end{equation}

As is clear from Eq.~(\ref{fr2}), the standard cosmological model of the current dark energy, described by the observed cosmological constant (\ref{cosmc}), is recovered by the extreme fine-tuning of the parameters in the 
UDW potential. Fine-tuning of the parameters of the effective (quintessence) scalar potential is the common property of {\it all} known models of the present dark energy.  Usually, one assumes that the mass of the quintessence scalar $\phi$ is of the order of the current Hubble scale, ie. $m\sim{\cal O}(10^{-33})$ eV. In our context it means that both parameters $(\m,u)$ are of the order one. However, in our UDW model, we can also assume that $\m \ll 1$ and
$u\geq 1$, so that the values of $\Lambda$ and $m$ decouple from each other.

\section{The $f(R)$ gravity corrections in our model}\label{sec:fgravity}

The difference between the cosmological constant and an $f(R)$ gravity model is described by the corrections beyond the leading term in Eq.~(\ref{fr2}),  which follow from the $f(R)$ gravity function. In our case, when using the parametric form of the $f$-function given by Eq.~(\ref{invs})  in the approximation (\ref{exp2}) and keeping the subleading terms, a tedious calculation gives
\be  \lb{subs}
  f(R) \approx e^{+\mu^{2} ( R / R_{0} ) }
    \left[ \lambda u ( u + 3 ) \mu^{4} \left( \frac{R}{R_{0}} \right)^{2} - \left( 1 + \frac{\mu^{2}}{u} \right) R
      + R_{0} \left( 1 - \frac{u + 4}{u} \mu^{2} \right) \right]~,
\ee
where we have ignored the higher-order terms in $R$ and $\m^2$, and
have introduced the background scalar curvature $R_0=-2\Lambda=-4\lambda\mu^2 u^2$.

We also verifed that the stability conditions (\ref{stab}) are satisfied for the values of $\abs{R}$ near and larger than the $\abs{R_0}$ in the regions I and II (as, e.g., in the Solar System and beyond it), i.e. $f'(R)<0$ and $f''(R)>0$, 
as long as $R$ is negative.

\begin{figure}[t]
 \centering
  \includegraphics[width=\hsize]{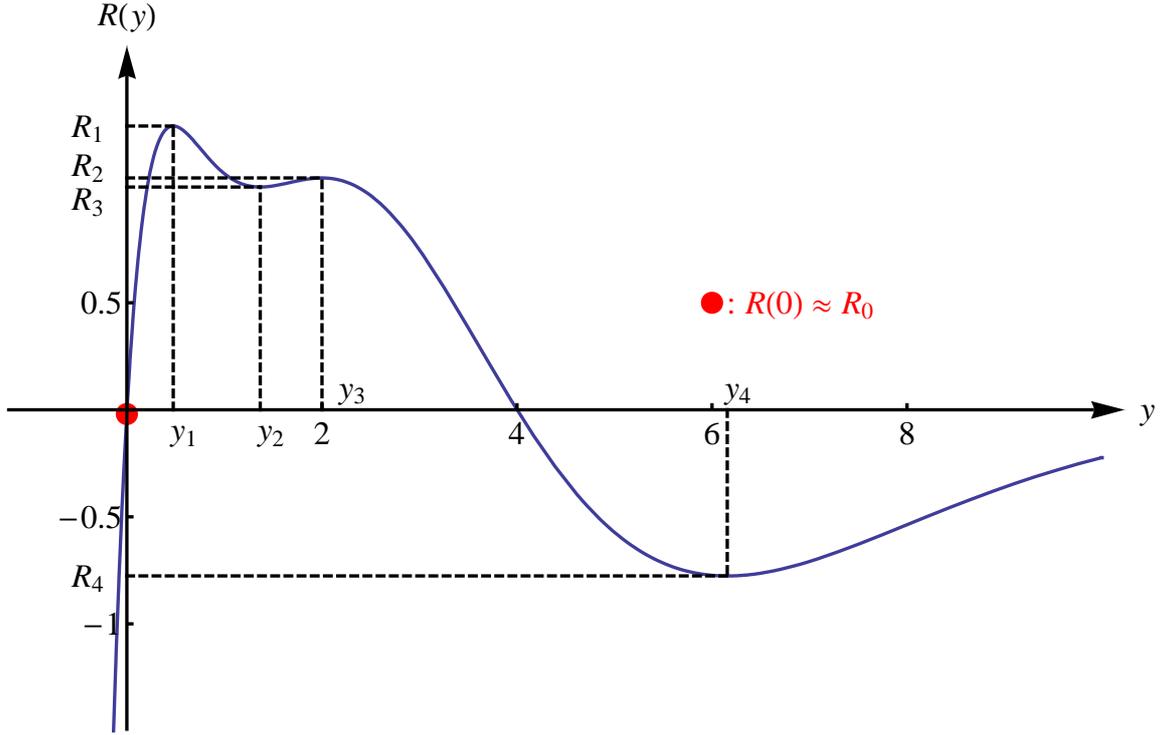}
      \caption{\small The profile of the function $R(y)/(2\lambda)$ for $u = 1$ and $\mu^{2} = 10^{-3}$.}
   \label{Rprof}
\end{figure}

Let us consider the regions I and II in Fig.~1 away from the dS minimum, with $|R|$ much larger than $|R_{0}|$,
in more detail. The behavior of the function $R(y)$ in the first line of Eq.~(\ref{invs}) after its renormalization by $\lambda$ is plotted in Fig.~\ref{Rprof}. 

For numerical calculations by the use of MATHEMATICA, we choose the values of the shape parameters as $u = 1$ 
and $\mu^{2} = 10^{-3}$, in agreement with our basic assumptions that $\mu$ is small and $u$ is large.

The function $R(y)$ has four extrema denoted by $R_{i}$ and $y_{i}$ with $i = 1,~2,~3,~4$. The $y_{3}$ and $y_{4}$ are in the vicinity of $2u$, and they are strongly dependent upon the value of $u$ but are slightly dependent upon the value of $\mu$. When $\mu \ll 1$,  $y_{1}$ and $y_{2}$ reach the values $0.6$ and $3.4$, respectively. When $u$
becomes large, $u\geq 10$, the values of  $y_{1}$ and $y_{2}$  are essentially independent upon the values of $u$ and $\mu$. For very large $\mu$ both $y_{1}$ and $y_{2}$ become imaginary and the extrema of the $R(y)$ for those $y$'s vanish.

The extremal values of $R_{1}$ and $R_{2}$ are approximately given by
\begin{equation} \lb{extr}
R_1 \approx 3.7 \lambda  u^{2}  \quad {\rm  and} \quad R_2  \approx  -1.3 \lambda u^{2} 
\end{equation}
whereas $R(0)\approx R_0$. As $y$ increases, the $R(y)$ reaches zero. Since the values of $y_{3}$ and $y_{4}$ strongly depend upon the value of $u$, both $R_{3}$ and $R_{4}$ are close to zero in the large $y$ region.
As is shown in Fig.~\ref{Rprof}, the $R(y)$ has extrema only in the region where $y$ is positive. When $y$ is negative, the $R(y)$ rapidly falls down to $-\infty$. We find the following approximations for $|R| \gg |R_{0}|$:
\begin{itemize}
  \item
    Region I: it appears that already for $y\approx -1$ we can take $|R| \gg |R_{0}|$. Then the higher order terms 
in $y$ inside the square brackets of the $R(y)$ can be ignored.  Hence, we can  approximate the $R(y)$ as
      \begin{equation}
        R(y) \approx - 2 \lambda e^{- y}
          \left[ y - 2 ( 2 u + 1 ) y + 2 u ( 2 u + 3 ) y - 4 u ( 2 u + \mu^{2} ) y + 2 \mu^{2} u ( 2 u + 1 ) \right]~.
      \end{equation}
  \item
    Region II: when $y$ is larger than the $y_{2}$, the $R(y)$ reaches zero and, therefore, the condition 
$|R| \gg |R_{0}|$ is not valid in this region. However, both  $|R_{1}|$ and $|R_{2}|$ are still significantly larger 
than the $|R_{0}|$. Hence, an expansion of the $R(y)$ around the $y_{2}$ can be valid too.
\end{itemize}

To the end of this Section, we calculate an expansion of the $f(R)$ function in the vicinity of $R_{1}\approx
3.7 \lambda u^{2}$ up to the second order in $(R-R_1)$.  It is the very extreme (non-physical) case far away from the dS vacuum, which helps us to make some qualitative conclusions in Sec.~5.  Given $u\geq 1$ and $\m \ll1$, we find 
\begin{equation} \label{allde}
  f(R_{1}) \approx - 1.1\lambda u^2~, \quad  f'(R_{1}) \approx 0.07~,
\quad  f''(R_{1}) \approx \frac{0.09}{\lambda u^{2}}~.
\end{equation}
It yields
\begin{equation}
  f(R) |_{R \approx R_{1}} \simeq \frac{0.09}{2 \lambda u^{2}} R^{2} - 0.25  R - 0.77\lambda u^2~~.
\end{equation}
At $R=R_{1}$ we find a violation of the classical stability  condition $f'(R)<0$.

\section{Conclusion}

We conclude that the Higgs-like quintessence scalar potential (\ref{udwp}) after fine-tuning of its parameters $(\lambda,\mu,u)$ is a viable proposal for the present dark energy. According to the proposal, our Universe is
in a meta-stable dS vacuum. This description is still valid for the average scalar curvatures $|R| \gg |R_0|$ (near
an observer) but breaks down for $R$ approaching $R_1$ (in the case of the Solar system, its average scalar
curvature $R_s$ obeys $|R_0| \ll |R_s| \ll |R_1|$). 

In our simple model the cosmological constant $\Lambda$ and the mass of the quintessence scalar $m$ decouple, being independent upon each other.

Our results also imply that the effective ``coupling constants" $(\Lambda,G_{\rm N},M)$ are very slowly dependent upon $R$ of the observer. Hence, our dynamical model of dark energy can be distinguished from the standard description (by $\Lambda)$ via a possible time (and/or space) dependence of the observed values of $(\Lambda,G_{\rm N},M)$ on cosmological scales.

\section*{Acknowledgements}

This work was supported by the Tokyo Metropolitan University and the World Premier International Research Center Initiative (WPI Initiative), MEXT, Japan. The authors thank Muhammad Usman for correspondence.

\end{document}